\documentclass[aps,preprint]{revtex4}
\usepackage{graphics}
\begin{document}

\preprint{PRL-draft-04-07-03}
\title[YBCO]{Search for quantum transducers between electromagnetic and gravitational 
radiation: A measurement of an upper limit on 
the transducer conversion efficiency of yttrium barium copper oxide}
\author{Raymond Y. Chiao\footnote{E-mail: chiao@physics.berkeley.edu}}
\affiliation{Department of Physics, University of California at Berkeley, CA 94720-7300}
\author{Walter J. Fitelson\footnote{E-mail: walt@ssl.berkeley.edu}}
\affiliation{University of California, Space Sciences Laboratory, Berkeley, CA
94720-7450}
\author{Achilles D. Speliotopoulos\footnote{E-mail: adspelio@uclink.berkeley.edu}}
\affiliation{Dept.~of Physics, University of California at Berkeley, CA 94720-7300}
\pacs{PACS number}

\begin{abstract}
A minimal coupling rule for the 
coupling of the electron spin to curved 
spacetime in general relativity suggests the possibility of a coupling between 
electromagnetic and gravitational radiation mediated by means of a quantum fluid.  
Thus quantum transducers between these two kinds of 
radiation fields might exist.  We report here on the first attempt at a Hertz-type experiment, 
in which a high-$\rm{T_c}$ superconductor (YBCO) was the sample material used as a 
possible quantum transducer to convert EM into GR microwaves, and a 
second piece of YBCO in a separate apparatus was used to 
back-convert GR into EM microwaves.  An upper limit on the conversion efficiency 
of YBCO was measured to be $1.6\times10^{-5}$ at liquid nitrogen temperature.
\end{abstract}

\volumeyear{year}
\volumenumber{number}
\issuenumber{number}
\eid{identifier}
\date[Date of draft: ]{April 7, 2003}
\received[Received date: ]{xxxx, 2003}
\maketitle

An interesting question that naturally arises at the border between 
general relativity (GR) and quantum mechanics (QM) is the
following: Are there novel experimental ways of studying
quantized fields coupled to curved spacetimes? ~This question has already
arisen in the context of the vacuum embedded in a curved spacetime \cite%
{DaviesBook}, and has already led to interesting theoretical suggestions 
of the existence of Hawking and Unruh radiations. 
Here we would like to extend this question to include experimental studies 
concerning the interaction between the ground state of a nonrelativistic quantum many-body system
with off-diagonal long-range order (ODLRO), i.e., a ``quantum fluid,'' viewed as the ``vacuum'' state of a
quantized field, and dynamically, but weakly, 
curved spacetimes, i.e., gravitional radiation fields \cite{ChiaoWheeler}. 
One implication of the theoretical analysis \cite{ChiaoWheeler} is that there might exist 
a quantum-fluid--mediated transduction 
process, in which electromagnetic (EM) radiation, such as microwaves, might be convertible 
into gravitational (GR) radiation, and vice versa. The size of the conversion efficiency of this 
process, however, is difficult to predict theoretically, and is best measured experimentally. 
We report here the first attempt to make such a measurement.

The electron, which possesses charge $e$, rest mass $m$, and spin $s=1/2$,
obeys the Dirac equation. The nonrelativistic, interacting, fermionic
many-body system, such as that in the quantum Hall fluid, should obey the
minimal-coupling rule which originates from the covariant-derivative
coupling of the Dirac electron to curved spacetime, viz.~(using the Einstein
summation convention), \cite{Weinberg}\cite{DaviesBook}
\begin{equation}
p_{\mu }\rightarrow p_{\mu }-eA_{\mu }-\frac{1}{2}\Sigma _{AB}\omega _{\mu
}^{AB}
\end{equation}%
where $p_{\mu }$ is the electron's four-momentum, $A_{\mu }$ is the
electromagnetic four-potential, $\Sigma _{AB}$ are the Dirac $\gamma $
matrices in curved spacetime with tetrad (or vierbein) $A,B$ indices, and $%
\omega _{\mu }^{AB}$ are the components of the spin connection%
\begin{equation}
\omega _{\mu }^{AB}=e^{A\nu }\nabla _{\mu }\left. e^{B}\right. _{\nu }
\end{equation}%
where $e^{A\nu }$ and $\left. e^{B}\right. _{\nu }$ are tetrad four-vectors,
which are sets of four orthogonal unit vectors of spacetime, such as those
corresponding to a local inertial frame.

Spacetime curvature directly affects the phase of the wavefunction, 
leading to fringe shifts of quantum-mechanical interference patterns within 
atom interferometers \cite{GRGessay}. Moreover, it is well known that 
the vector potential $A_{\mu }$ will also lead to a quantum interference effect, in 
which the gauge-invariant Aharonov-Bohm phase becomes observable. Similarly, 
the spin connection $\omega _{\mu }^{AB}$, in its Abelian holonomy, should 
also lead to a quantum interference effect, in which the gauge-invariant 
Berry phase \cite{Chiao-Wu} becomes observable. The following Berry phase 
picture of a spin coupled to curved spacetime leads to an intuitive way of 
understanding why there could exist a coupling between a classical GR wave 
and a classical EM wave mediated by a quantum fluid with charge and spin, such  
as the quantum Hall fluid.

Due to its gyroscopic nature, the spin vector of an electron undergoes \emph{%
parallel transport} during the passage of a GR wave. The spin of the
electron is constrained to lie inside the space-like submanifold of curved
spacetime. This is due to the fact that we can always transform to a
co-moving frame, such that the electron is at rest at the origin of this
frame. In this frame, the spin of the electron must be purely a space-like
vector with no time-like component. This imposes an important $constraint$
on the motion of the electron's spin, such that whenever the space-like
submanifold of spacetime is disturbed by the passage of a gravitational
wave, the spin must remain at all times $perpendicular$ to the local time
axis. If the spin vector is constrained to follow a conical trajectory
during the passage of the gravitational wave, the electron picks up a Berry
phase proportional to the solid angle subtended by this conical trajectory
after one period of the GR wave.

In a manner similar to the persistent currents induced by the Berry phase in
systems with ODLRO \cite{Lyanda-Geller}, such a Berry
phase induces a macroscopic electrical current in the quantum fluid, such 
as the quantum Hall fluid, which is in a
macroscopically coherent ground state \cite{Girvin}. This current generates
an EM wave. Thus a GR wave can be converted into an EM wave. By reciprocity,
the time-reversed process of the conversion from an EM wave to a GR wave
must also be possible.

In the nonrelativistic limit, the four-component Dirac spinor is reduced to
a two-component spinor. While the precise form of the nonrelativistic
Hamiltonian is not known for the many-body system in a weakly curved
spacetime consisting of electrons in a strong magnetic field, one expects 
that it will have the form%
\begin{equation}
H=\frac{1}{2m}\left( p_{i}-eA_{i}-\frac{1}{2}\sigma_{ab} {\Omega}%
_{i}^{ab}\right) ^{2}+V
\end{equation}%
where $i$ is a spatial index, $a,b$ are spatial tetrad incides, $\sigma_{ab}$
is a two-by-two matrix-valued tensor representing the spin, and $\sigma_{ab}{%
\Omega}_{i}^{ab}$ is the nonrelativistic form of $\Sigma _{AB}\omega _{\mu
}^{AB}$. Here $H$ and $V$ are two-by-two matrix operators on the
two-component spinor electron wavefunction in the nonrelativistic limit. The
potential energy $V$ includes the Coulomb interactions between the electrons
in the quantum Hall fluid. This nonrelativistic Hamiltonian has the form%
\begin{equation}
H=\frac{1}{2m}\left( \mathbf{p}-\mathbf{a}-\mathbf{b}\right) ^{2}+V\mbox{ ,}
\end{equation}%
where the particle index, the spin, and the tetrad indices have all been
suppressed. Upon expanding the square, it follows that for a quantum Hall
fluid of uniform density, there exists a cross-coupling or interaction
Hamiltonian term of the form%
\begin{equation}
H_{int}\sim \mathbf{a}\cdot \mathbf{b}\mbox{ ,}  \label{cross-coupling}
\end{equation}%
which couples the electromagnetic $\mathbf{a}$ field to the gravitational $%
\mathbf{b}$~field. In the case of time-varying fields, $\mathbf{a}(t)$ and $%
\mathbf{b}(t)$ represent EM and GR radiation, respectively.

\begin{figure}
\centerline{\includegraphics{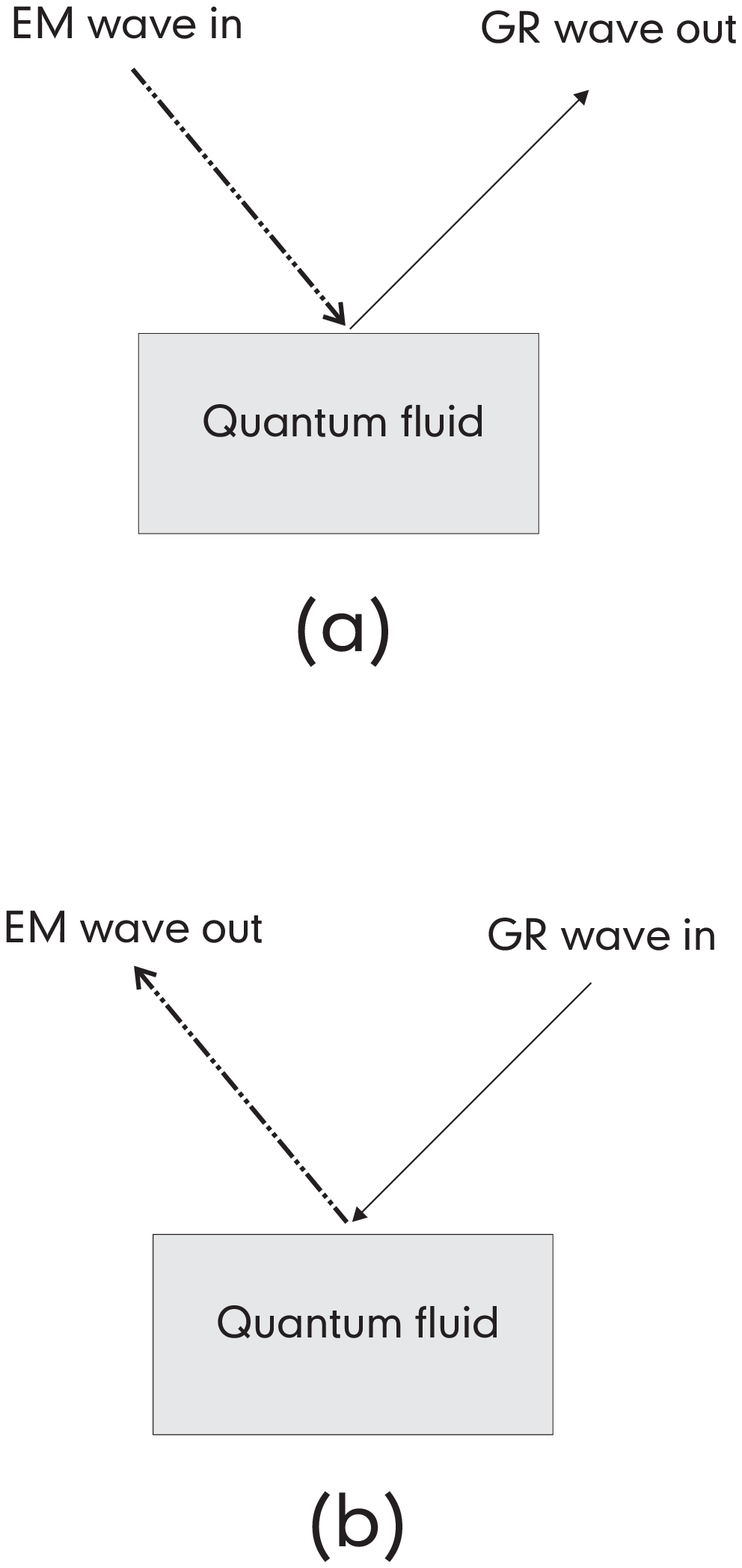}}
\caption{Quantum transducer...}
\label{Fresnel}
\end{figure}

In first-order perturbation theory, the quantum adiabatic theorem predicts
that there will arise the cross-coupling energy between the two radiation
fields mediated by this quantum fluid 
\begin{equation}
\Delta E \sim \langle \Psi_0|\mathbf{a}\cdot \mathbf{b}|\Psi_0 \rangle
\end{equation}
where $|\Psi_0 \rangle$ is the unperturbed ground state of the system. For
the adiabatic theorem to hold, there must exist an energy gap $E_{gap}$
(e.g., the quantum Hall energy gap) separating the ground state from all
excited states, in conjunction with a time variation of the radiation fields
which must be slow compared to the gap time $\hbar/E_{gap}$. This suggests
that under these conditions, there might exist an interconversion process
between these two kinds of classical radiation fields mediated by this
quantum fluid, as indicated in Fig.\ref{Fresnel}.

The question immediately arises: EM radiation is fundamentally a spin 1
(photon) field, but GR radiation is fundamentally a spin 2 (graviton) field.
How is it possible to convert one kind of radiation into the other, and not
violate the conservation of angular momentum? ~The answer: The EM wave
converts to the GR wave \emph{through a medium}. Specifically, in the case 
of the quantum Hall fluid, the
medium of conversion consists of a strong DC magnetic field applied to a
system of electrons. This system possesses an axis of symmetry pointing
along the magnetic field direction, and therefore transforms like a spin 1
object. When coupled to a spin 1 (circularly polarized) EM radiation field,
the total system can in principle produce a spin 2 (circularly polarized) GR
radiation field, by the addition of angular momentum. However, it remains an
open question as to how strong this interconversion process is between EM
and GR radiation. Most importantly, the size of the conversion efficiency of
this transduction process needs to be determined by experiment.

We can see more clearly the physical significance of the interaction
Hamiltonian $H_{int}\sim \mathbf{a}\cdot \mathbf{b}$ once we convert it into
second quantized form and express it in terms of the creation and
annihilation operators for the positive frequency parts of the\ two kinds of
radiation fields, as in the theory of quantum optics, so that in the
rotating-wave approximation%
\begin{equation}
H_{int}\sim a^{\dagger }b+b^{\dagger }a\mbox{ ,}
\end{equation}%
where the annihilation operator $a$ and the creation operator $a^{\dagger }$
of the single classical mode of the plane-wave EM radiation field
corresponding the $\mathbf{a}$ term, obey the commutation relation $%
[a,a^{\dagger }]=1$, and where the annihilation operator $b$ and the
creation operator $b^{\dagger }$ of the single classical mode of the
plane-wave GR radiation field corresponding to the $\mathbf{b}$ term, obey
the commutation relation $[b,b^{\dagger }]=1$. The first term $a^{\dagger }b$ then
corresponds to the process in which a graviton is annihilated and a photon
is created inside the quantum fluid, and similarly the second term $%
b^{\dagger }a$ corresponds to the reciprocal process, in which a photon is
annihilated and a graviton is created inside the quantum fluid \cite{crude}.

One may ask whether there exists $any$ difference in the response of quantum
fluids to tidal fields in gravitational radiation, and the response of
classical matter, such as the lattice of ions in a superconductor, for
example, to such fields. The essential difference between quantum fluids and
classical matter is the presence or absence of macroscopic quantum
interference. In classical matter, such as in the lattice of ions of a
superconductor, decoherence arising from the environment destroys any such
quantum interference. Hence, the response of quantum fluids and of classical
matter to these fields will therefore differ from each other \cite%
{ChiaoWheeler}.

In the case of superconductors, Cooper pairs of electrons possess a
macroscopic phase coherence, which can lead to an Aharonov-Bohm-type
interference absent in the ionic lattice. Similarly, in the quantum Hall
fluid, the electrons will also possess macroscopic phase coherence \cite%
{Girvin} which can lead to Berry-phase-type interference absent in the
lattice. Furthermore, there exist ferromagnetic superfluids with intrinsic
spin in which an ionic lattice is completely absent, such in
superfluid helium 3 \cite{Osheroff}and spin-polarized atomic BECs 
\cite{CornellWieman}. 
In such ferromagnetic quantum fluids, there exists no
ionic lattice to give rise to any classical response which could prevent a
quantum response to tidal gravitational radiation fields. The
Berry-phase-induced response of the ferromagnetic superfluid arises from the
spin connection (see the above minimal-coupling rule, which can be
generalized from an electron spin to a nuclear spin), and leads to a purely
quantum response to this radiation. The Berry phase induces time-varying
macroscopic quantum flows in this ferromagnetic ODLRO system \cite%
{Lyanda-Geller} which transports time-varying orientations of the nuclear
magnetic moments, and thus generates EM waves. This ferromagnetic superfluid
can therefore also in principle interconvert GR into EM radiation, and vice
versa, in a manner similar to the case discussed above for the ferromagnetic
quantum Hall fluid. Thus there may be more than one kind of quantum fluid
which can serve as a transducer between EM and GR waves.

Like superfluids, the quantum Hall fluid is an example of a quantum fluid
which differs from a classical fluid in its current-current correlation
function in the presence of GR waves. In particular, GR waves can induce a
transition of the quantum Hall fluid out of its ground state $only$ by
exciting a quantized, collective excitation, such as the vortex-like $\frac{1%
}{3}e$ quasi-particle, across the quantum Hall energy gap. This collective
excitation would involve the correlated motions of a macroscopic number of
electrons in this coherent quantum system. Hence the quantum Hall fluid,
like the other quantum fluids, should be effectively incompressible and
dissipationless, and is thus a good candidate for a quantum antenna and
transducer.

There exist other situations in which a minimal-coupling rule similar to the
one above, arises for $scalar$ quantum fields in curved spacetime. DeWitt 
\cite{DeWitt} suggested~in 1966 such a coupling in the case of
superconductors. Speliotopoulos noted in 1995 \cite{Speliotopoulos1995} that
a cross-coupling term of the form $H_{int}\sim \mathbf{a}\cdot \mathbf{b}$
arose in the long-wavelength approximation of a certain quantum Hamiltonian
derived from the geodesic deviation equations of motion using the
transverse-traceless gauge for GR waves.

Two of us (ADS adn RYC) have been working on the problem of the coupling of a
scalar quantum field to curved spacetime in a general laboratory frame,
which avoids the use of the long-wavelength approximation \cite{GLF}. In
general relativity, there exists no global time coordinate that
can apply throughout a large system for nonstationary metrics, such
as those associated with gravitational radiation, since the local time axis varies
from place to place in the system. It is therefore necessary to set up
operationally a general laboratory frame by which an observer can measure
the motion of slowly moving test particles in the presence of weak,
time-varying gravitational radiation fields.

For either a classical or quantum test particle, the result is that its mass 
$m$ should enter into the Hamiltonian through the replacement of $\mathbf{p}%
-e\mathbf{A}$ by $\mathbf{p}-e\mathbf{A}-m\mathbf{N}$, where $\mathbf{N}$ is
the small, local tidal velocity field induced by gravitational radiation on
a test particle located at $X_a$ relative to the observer at the origin
(i.e., the center of mass) of this frame, where, for the small deviations $%
h_{ab}$ of the metric from that of flat spacetime, 
\begin{equation}
N_a=\frac{1}{2}\int_{0}^{X_a} \frac{\partial h_{ab}}{\partial t} dX^{b}.
\end{equation}
Due to the quadrupolar nature of gravitational tidal fields, the velocity
field $\mathbf{N}$ for a plane wave grows linearly in magnitude with the
distance of the test particle as seen by the observer located at the center
of mass of the system. Therefore, in order to recover the standard result of
classical GR that only $tidal$ gravitational fields enter into the coupling
of radiation and matter, one expects in general that a new characteristic
length scale $L$ corresponding to the typical size of the distance $X_a$
separating the test particle from the observer, must enter into the
determination of the coupling constant between radiation and matter. For
example, $L$ can be the typical size of the detection apparatus (e.g., the
length of the arms of the Michelson interferometer used in LIGO), or of the
transverse Gaussian wave packet size of the gravitational radiation. 
For the case of superconductors, treating Cooper pairs of
electrons as bosons, we would expect the above arguments would carry over
with the charge $e$ replaced by $2e$ and the mass $m$ replaced by $2m$.

Motivated by the above theoretical considerations, we performed an
experiment using a high $T_{c}$ superconductor, yttrium barium copper oxide
(YBCO), as one such possible quantum transducer, in a first attempt to
observe the predicted quantum transduction process from EM to GR waves, and
vice versa. We chose YBCO because it allowed us to use liquid nitrogen as
the cryogenic fluid for cooling the sample down below $T_{c}=90$ K to
achieve macroscopic quantum coherence, which is much simpler to do than to use 
liquid helium. Although we did not observe a detectable conversion signal in
this first experiment, we did establish an upper bound on the transducer
conversion efficiency of YBCO, and the techniques we used in this experiment
could prove to be useful in future experiments.

The idea of the experiment was as follows: Use a first YBCO sample to
convert EM into the GR radiation by shining microwaves onto it, and use a
second sample to back-convert the GR radiation generated in the far field by
the first sample back into EM radiation of the original frequency. In this
way, GR radiation could be generated by the first YBCO sample as the \emph{%
source} of such radiation inside a first closed metallic container, and GR
radiation could be detected by the second sample as the \emph{receiver} of
such radiation inside a second closed metallic container, in a Hertz-type
experiment.

The electromagnetic coupling between the two halves of the apparatus
containing the two YBCO samples, called the ``Emitter'' and the
``Receiver,'' respectively, could be prevented by means of two Faraday
cages, i.e., the two closed normal metallic cans which completely surrounded the
two samples and their associated microwave equipment. See Fig.\ref%
{Rough-Schematic}. The Faraday cages consisted of two empty one-gallon paint
cans with snugly fitting cover lids, whose inside walls, cover lids, and can
bottoms, were lined on their interiors with a microwave-absorbing foam-like
material (Eccosorb AN70), so that any microwaves incident upon these walls
were absorbed. Thus multiply-reflected EM microwave radiation within the
cans could thereby be effectively eliminated.

The electromagnetic coupling between the two cans with their cover lids on,
was measured to be extremely small (see below). Since the Faraday cages were
made out of normal metals, and the Eccosorb materials were also not composed
of any macroscopically coherent quantum matter, these shielding materials
should have been essentially transparent to GR radiation. Therefore, we
would expect that GR radiation should have been able to pass through from
the source can to the receiver can without much attenuation.

\begin{figure}
\centerline{\includegraphics{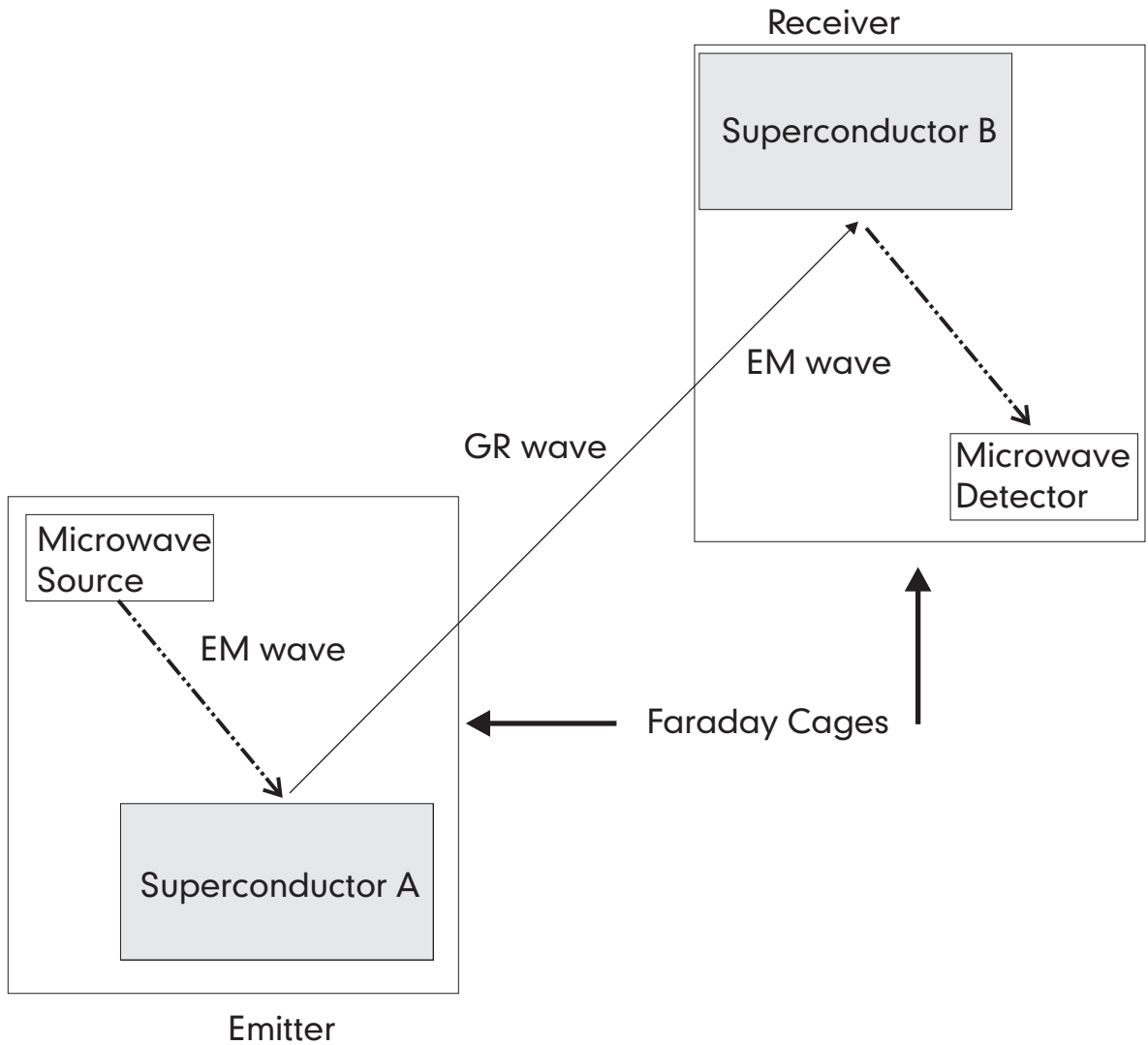}}
\caption{Simplified schematic of a Hertz-type experiment...}
\label{Rough-Schematic}
\end{figure}

A simplified schematic outlining the Hertz-type experiment is shown in Fig.%
\ref{Rough-Schematic}, in which gravitational radiation at 12 GHz could be
emitted and received using two superconductors. The ``Microwave Source'' in
this Figure generated electromagnetic radiation at 12 GHz (``EM wave''),
which was directed onto Superconductor A (the first piece of YBCO) immersed
in liquid nitrogen, and would be converted upon reflection into
gravitational radiation (``GR wave''). \ 

The GR wave, but not the EM wave, could pass through the ``Faraday Cages.''
In the far field of Superconductor A, Superconductor B (a second piece of
YBCO), also immersed in liquid nitrogen, could reconvert upon reflection the
GR wave back into an EM wave at 12 GHz, which could then be detected by the
``Microwave Detector.''

To prevent transitions out of a macroscopically 
coherent quantum state in YBCO, the
frequency of the microwaves was chosen to be well below the superconducting
gap frequency of YBCO. In order to satisfy this requirement, we chose for
our experiment the convenient microwave frequency of 12 GHz (or a wavelength
of 2.5 cm), which is three orders of magnitude less than the typical gap frequency of
YBCO.

Since the predicted conversion process is fundamentally quantum mechanical
in nature, the signal would be predicted to disappear if either of the two
samples were to be warmed up above the superconducting transition
temperature. Hence the signal at the microwave detector should disappear
once either superconductor was warmed up above its transition temperature,
i.e., after the liquid nitrogen boiled away in either dewar containing the
YBCO samples.

It should be emphasized that the predicted quantum transducer conversion
process involves a linear relationship between the amplitudes of the two
kinds of radiation fields (EM and GR), since we are considering the linear
response of the first sample to the incident EM wave during its generation
of the outgoing GR wave, and also the linear response of the second sample
to the incident GR wave during its generation of the outgoing EM wave.
Time-reversal symmetry, which has been observed to be obeyed by EM and GR
interactions at low energies for classical fields, would lead us to expect
that these two transducer conversion processes obey the principle of
reciprocity, so that the reverse process should have an efficiency equal to
that of the forward process.

Thus, assuming that the two samples are essentially identical, we expect that the
overall power conversion efficiency of this Hertz-type experiment $\eta
_{Hertz}$ should be 
\begin{equation}
\eta _{Hertz}=\eta _{EM\rightarrow GR}\cdot \eta _{GR\rightarrow EM}=\eta
^{2}  \label{eta}
\end{equation}%
where $\eta _{EM\rightarrow GR}$ is the EM-to-GR power conversion efficiency
by the first sample, and $\eta _{GR\rightarrow EM}$ is the GR-to-EM power
conversion efficiency of the second sample. If the two samples are closely
similar to each other, we expect that $\eta _{EM\rightarrow GR}=$ $\eta
_{GR\rightarrow EM}= \eta $, where $\eta $ is the transducer power
conversion efficiency of identical samples. 

In the case of the quantum Hall fluid considered earlier, the medium would
have a strong magnetic field applied to it, so that the conservation of
total angular momentum during the conversion process between the spin-1 EM
field and the spin-2 GR field, could be satisfied by means of the angular
momentum exchange between the fields and the anisotropic quantum Hall
medium. Here, however, our isotropic, compressed-powder YBCO medium did not
have a magnetic field applied to it, so that it
was necessary to satisfy the conservation of angular momentum in another
way: One must first convert the EM field into an angular-momentum 2,
quadrupolar, far-field radiation pattern.

This was accomplished by means of a T-shaped electromagnetic antenna, which
generated in the far field an quadrupolar EM field pattern that matched that
of the quadrupolar GR radiation field pattern. In order to generate a
quadrupolar EM radiation field, it is necessary to use an antenna with
structure possessing an even-parity symmetry. This was implemented by
soldering onto the central conductor of a SMA coaxial cable a
one-wavelength-long wire extending symmetrically on either side of the
central conductor in opposite directions, in the form of a T-shaped antenna
(see Fig.3).

A one-inch cube aluminum block assembly was placed at approximately a
quarter of a wavelength behind the ``T,'' so as to reflect the antenna
radiation pattern into the forwards direction, and also to impedance-match
the antenna to free space. The aluminum block assembly consisting of two
machined aluminum half-blocks which could be clamped tightly together to fig
snugly onto the outer conductor of the SMA coaxial cable, so as to make a
good ohmic contact with it. The joint between the two aluminum half-blocks
was oriented parallel to the bar of the ``T.'' Thus the block formed a good
ground plane for the antenna. The resonance frequency of this T-antenna
assembly was tuned to be 12 GHz, and its $Q$ was measured to be about 10,
using a network analyzer (Hewlett Packard model HP8720A).

Measurements of the radiative coupling between two such T antennas placed
directly facing each other at a fixed distance, while scanning their relative
azimuthal angle, showed that extinction between the antennas occured at a
relative azimuthal angle of 45$^{\circ }$ between the two ``T''s, rather
than at the usual 90$^{\circ }$ angle expected for crossed dipolar antennas.
Furthermore, we observed that at a mutual orientation of 90$^{\circ }$
between the two T antennas (i.e., when the two ``T''s were crossed with
respect to each other), a $maximum$ in the coupling between the antennas, in
contrast to the $minimum$ expected in the coupling between two crossed
linear dipole antennas. This indicates that our T antennas were indeed
functioning as quadrupole antennas. Thus, they would generate a quadrupolar
pattern of EM radiation fields in the far field, which should be homologous
to that of GR radiation.

For generating the 12 GHz microwave beam of EM radiation, which we used for
shining a beam of quadrupolar radiation on the first YBCO sample, we started
with a 6 GHz ``brick'' oscillator (Frequency West model MS-54M-09), with an
output power level of 13 dBm at 6 GHz. This 6 GHz signal was amplified, and
then doubled in a second harmonic mixer (MITEQ model MX2V080160), in order
to produce a 12 GHz microwave beam with a power level of 7 dBm. The 12 GHz
microwaves was fed into the T antenna that shined a quadrupolar-pattern beam
of EM radiation at 12 GHz onto the first YBCO sample immersed in a liquid
nitrogen dewar inside the source can. The sample was oriented so as to
generate upon reflection a 12 GHz GR radiation beam directed towards the
second YBCO sample along a line of sight inside the receiver can (see Fig.3).

The receiver can contained the second YBCO sample inside a liquid nitrogen
dewar, oriented so as to receive the beam of GR, and back-convert it into a
beam of EM radiation, which was directed upon reflection towards a second T
antenna. A low-noise preamp (Astrotel model PMJ-LNB KU, used for receiving
12 GHz microwave satellite communications), which had a noise figure of 0.6
dB, was used as the first-stage amplifier of the received signal. This noise 
figure determined the overall sensitivity of the measurement. This
front-end LNB (Low-Noise Block) assembly, besides having a low-noise preamp,
also contained inside it an internal mixer that down-converted the amplified
12 GHz signal into a standard 1 GHz intermediate frequency (IF) band. We
then fed this IF signal into a commercial satellite signal level meter
(Channel Master model 1005IFD), which both served as the DC power supply for
the LNB assembly by supplying a DC voltage back through the center conductor
of a F-style IF coax cable into the LNB assembly, and also provided
amplification of the IF signal. Its output was then fed into a spectrum
analyzer (Hewlett-Packard model 8559A).

In order for the YBCO samples (1 inch diameter, 1/4 inch thick pieces of
high-density YBCO from Arbor Scientific) 
to become superconducting, we cooled these samples to 77K
by immersing them in liquid nitrogen. The dewars needed for holding this
cryogenic fluid together with the YBCO samples consisted of a stack of
styrofoam cups; the dead air space between the cups, which were glued
together at their upper lips, served as good thermal insulation.

The samples were epoxied in a vertical orientation into a slot in a
styrofoam piece which fit snugly into the bottom of the top cup of the
stack, and the cups also fit snugly into a hole in the top layer of Eccosorb
foam pieces placed at the bottom of the can. Also, since styrofoam was
transparent to microwave radiation, these cup stacks also served as
convenient $dielectric$ dewars for holding the YBCO samples in liquid
nitrogen. At the beginning of a run, we would pour into these cups liquid
nitrogen, which would last about a hour before it boiled away. The
temperatures of the samples were monitored by means of thermocouples
attached to the back of the samples.

We show in part (a) of Fig.\ref{Data2} data showing the IF\ spectrum
analyzer output of the signal from the receiver can with the cover lids $off$
both the source can and the receiver can, which allowed a small leakage
signal to be coupled between the two cans (to test whether the entire system
was working properly), and in part (b), data with covers lids $on$ both
cans. Both YBCO samples were immersed in liquid nitrogen for both (a) and
(b). The data in (b) show that the Eccosorb-lined Faraday cages were very
effective in screening out any electromagnetic pickup. However, there is no
detectable signal above the noise that would indicate any detectable
coupling due to the quantum transducer conversion between EM and GR waves.
Before taking these data, we tested \emph{in situ} that when they immersed
in liquid nitrogen, the YBCO samples were indeed superconducting 
by the observation of a repulsion away from the YBCO of a small permanent
magnet hung as the bob of a pendulum by means of a string near the samples.

The sensitivity of the source-receiver system was calibrated in a separate
experiment, in which we replaced the two T antennas by a low-loss cable
directly connecting the source to the receiver, in series with 70 dB of
calibrated attenuation. We could then measure the size of this directly
coupled 12 GHz electromagnetic signal on the spectrum analyzer with respect
to the noise rise, which served as a convenient measure of the minimum
detectable signal strength. In the resulting spectrum, which was similar to
that shown in Fig.\ref{Data2}(a), we observed a $-77$ dBm central peak at 12
GHz, which was 25 dB above the noise rise. This implies that we could have
seen a signal of $-102$ dBm of transducer-coupled radiation with a
signal-to-noise ratio of about unity. Assuming that the T antennas were
perfectly efficient in coupling to the YBCO samples, from the data shown in
Fig.\ref{Data2} we would infer that the observed efficiency $\eta_{Hertz}$
was less than 95 dB, and therefore from Eq.(\ref{eta}), that the quantum
transducer efficiency $\eta$ was less than 48 dB, i.e., $\eta<1.6%
\times10^{-5}$.

Why did we even bother performing this transducer experiment, when we knew
that Faraday cages were essentially perfect shields, and therefore that
there seemingly should have been no coupling at all between the two cans?
The first answer: Even classically, one expects a $nonzero$ coupling between
the cans due to the fact that accelerated electrons produce a $nonvanishing$
amount of GR radiation, since each electron possesses a mass $m$, as well as
a charge $e$. Therefore, whenever an electron's charge undergoes
acceleration, so will its mass. Relativistic causality therefore
necessitates that changes in the gravitational field of an electron in the
radiation zone due to its acceleration must be retarded by the speed of
light, just like those of the electromagnetic field in the radiation zone. This
implies that there must exist a transducer power conversion efficiency of at
least $Gm^{2}\cdot 4\pi \varepsilon _{0}/e^{2}=2.4\times 10^{-43}$, based on
a naive classical picture in which each individual electron possesses a
deterministic, Newtonian trajectory. Thus even in principle, the Faraday
cages could not have provided a perfect shielding between the two cans.
However, if this classical picture had been correct, there would have been
no hope of actually observing this conversion process, based on the
sensitivity of existing experimental techniques described
above.

The second answer: Superconductivity is fundamentally a quantum mechanical
phenomenon. Due to the macroscopic coherence of the ground state with ODLRO,
and the existence of a non-zero energy gap, there may exist quantum
many-body enhancements to this classical conversion efficiency. In addition
to these enhancements, there must exist additional enhancements due to the
dependence of the coupling on the size of the sample $L$, 
in order to account correctly for the $tidal$ nature of GR
waves \cite{GLF}.

The third answer: The ground state of a superconductor, which possesses spontaneous symmetry
breaking, and therefore ODLRO, is very similar to
that of the physical vacuum, which is believed also to possess spontanous
symmetry breaking through the Higgs mechanism. In this sense, therefore, the
vacuum is ``superconducting.'' The question thus arises: How does such a
broken-symmetry ground, or ``vacuum,'' state interact with a dynamically
changing spacetime, such as that associated with a GR wave? More generally:
How do we embed quantum fields in dynamically curved spacetimes? We believe
that this question has never been explored before $experimentally$.

How then do we account for the lack of any observable quantum transducer
conversion in our experiment? There are several possible reasons, the most
important ones probably having to do with the material properties of the
YBCO samples. One such possible reason is the earlier observations of
unexplained residual microwave and far-infrared losses (of the order of 10$%
^{-5}$ ohms per square at 10 GHz) in YBCO and other high T$_{\mathrm{c}}$
superconductors, which are independent of temperature and have a
frequency-squared dependence \cite{Miller} which may be due to the fact that
YBCO is a $D$-wave superconductor \cite{Tinkham}. In $D$-wave
superconductors, there exists a four-fold symmetry of nodal lines along
which the\ BCS gap vanishes \cite{Davis}, where the microwave attenuation may
become large. Thus $D$-wave superconductors are quite unlike the classic,
low-temperature $S$-wave superconductors with respect to their microwave
losses. Since one of conditions for a good coupling of a quantum antenna and
transducer to the GR wave sector is extremely low dissipative losses, the
choice of YBCO as the material medium for the Hertz-type experiment may not
have been a good one.

\section{Acknowledgments}

We thank Dung-Hai Lee, Jon-Magne Leinaas, Robert Littlejohn, Joel Moore,
Richard Packard, Paul Richards, Daniel Solli, Achilles Speliotopoulos, Neal
Snyderman, and Sandy Weinreb for stimulating discussions. This work was
supported in part by the ONR.

\pagebreak

\section{Figure Captions}

Figure 1: Quantum transducer between electromagnetic (EM) and gravitational
(GR) radiation, consisting of a quantum fluid, such as the quantum Hall
fluid, which possesses charge and spin. The minimal-coupling rule for an
electron coupled to curved spacetime via its charge and spin, results in two
processes. In (a) an EM plane wave is converted upon reflection from the
quantum fluid into a GR plane wave; in (b), which is the reciprocal or
time-reversed process, a GR plane wave is converted upon reflection from the
quantum fluid into an EM plane wave.

Figure 2: Simplified schematic of a Hertz-type experiment, in which
gravitational radiation at 12 GHz could be emitted and received using two
superconductors. The ``Microwave Source'' generated electromagnetic
radiation at 12 GHz (``EM wave''), which impinged on Superconductor A, could
be converted upon reflection into gravitational radiation (``GR wave''). The
GR wave, but not the EM wave, could pass through the ``Faraday Cages.'' In
the far field of Superconductor A, Superconductor B could reconvert upon
reflection the GR wave back into an EM wave at 12 GHz, which could then be
detected by the ``Microwave Detector.'' 

Figure 3: The T-antenna (expanded view on the left) used as antennas inside
the ``Source Can'' and the ``Receiver Can.'' The YBCO samples were oriented
so that a GR microwave beam could be directed from one YBCO sample to the
other along a straight line of sight.

Figure 4: Data from the Hertz-type gravity-wave experiment using YBCO
superconductors as transducers between EM and GR radiation. In (a), the
cover lids were off both the source and the receiver cans, so that a small
leakage signal (the central spike) could serve to test the system. In (b),
both cover lids were on the cans, but no detectable signal of coupling
between the cans could be seen above the noise. Both YBCO samples were
immersed in liquid nitrogen for these data.

\pagebreak
\pagebreak
\pagebreak
\begin{figure}
\centerline{\includegraphics{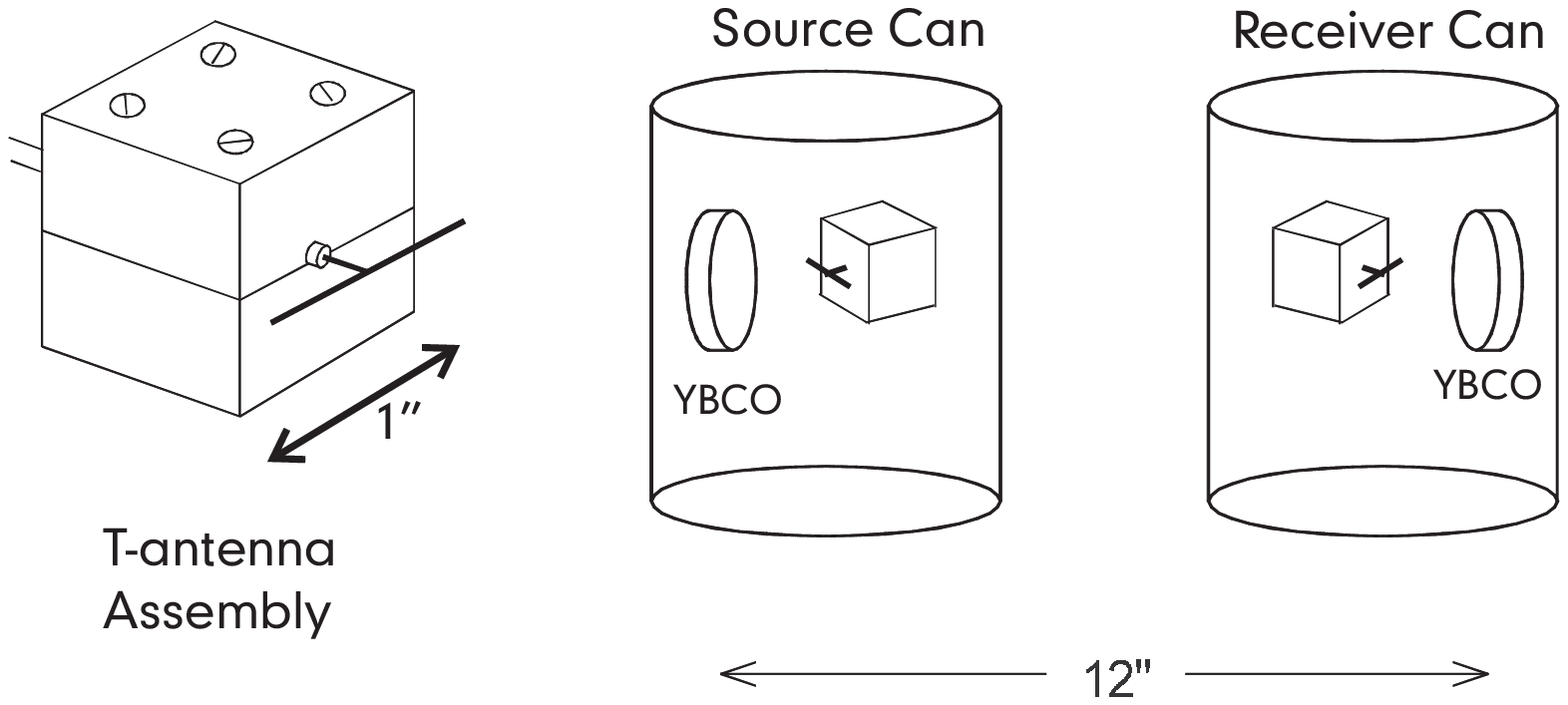}}
\caption{The T-antenna...}
\label{T-Antennas}
\end{figure}
\pagebreak
\begin{figure}
\centerline{\includegraphics{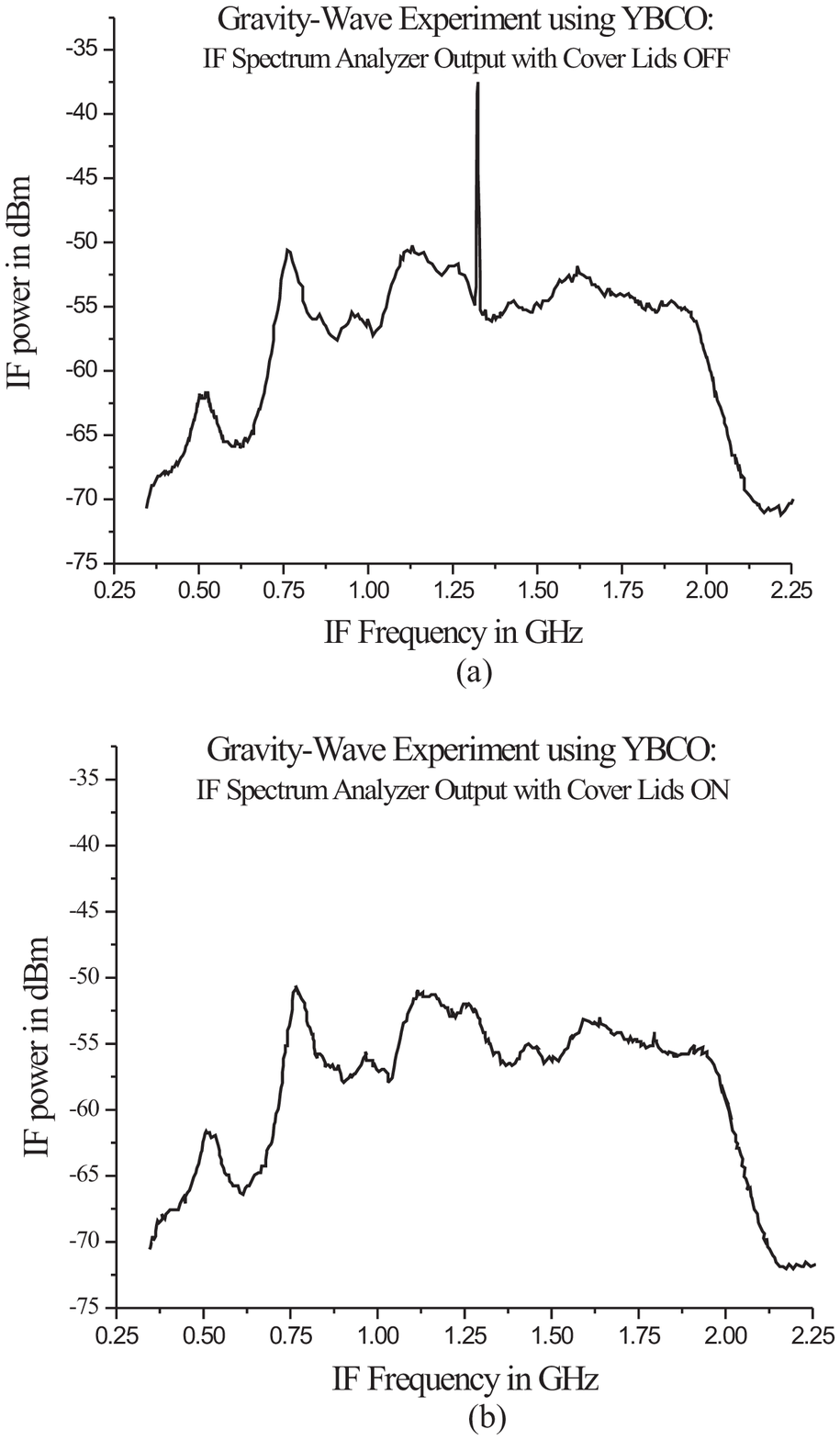}}
\caption{Data from the Hertz-type...}
\label{Data2}
\end{figure}

\end{document}